\RequirePackage{lineno}
\documentclass[aps,prl,twocolumn,showpacs,superscriptaddress,groupedaddress]{revtex4}  % for review and submission 

\usepackage{graphicx}  % needed for figures

\usepackage{dcolumn}   % needed for some tables
\usepackage{bm}        % for math
\usepackage{amssymb}   % for math
\usepackage{amsmath}   % for math
\usepackage{xspace}

\RequirePackage{color}\definecolor{RED}{rgb}{1,0,0}\definecolor{BLUE}{rgb}{0,0,1} \definecolor{PURPLE}{rgb}{1,0,1}%DIF PREAMBLE

\usepackage{color}
\newcommand{\CHECK}[1]{\textbf{\color{red}[#1]}\xspace}
\newcommand{\chk}[1]{\CHECK{CHECK THIS!}}

\newcommand{\fb}{\ensuremath{\mathrm{fb}^{-1}}}
\newcommand{\Eslash}{\mbox{$ E \kern-0.6em\slash$                }}
\newcommand{\pslash}{\mbox{$ p \kern-0.45em\slash$                }}
\newcommand{\etmiss}{\mbox{$\Eslash_T\!$                        }}

\newcommand{\ttbar}{\mbox{$t{\bar t}~$}}

\newcommand{\la}{\langle}
\newcommand{\ra}{\rangle}

\newcommand{\wc}{$W+c$-jet\xspace}
\newcommand{\wb}{$W+b$-jet\xspace}

\def\MET{{\mbox{$E\kern-0.57em\raise0.19ex\hbox{/}_{T}$}}}
\def\met{{\mbox{$E\kern-0.57em\raise0.19ex\hbox{/}_{T}$}}}

\def\ppbar{$p\bar{p}$}

\def\lmet{$WH\rightarrow \ell\kern-0.45em\raise0.19ex\hbox{/} \nu b\bar{b}$}

\begin{document}

\hspace{5.2in} \mbox{FERMILAB-PUB-14-525-E}

%\leftline{\em PLB draft, Version 2.4}
%\today
%\leftline{Primary Authors: Dmitry Bandurin, Oleksandr Borysov,}
%\leftline{Olga Gogota and Bj\"orn Penning}

%\rightline{\em ~~~~~~~~~~~~~~~~~~~ D\O\ INTERNAL DOCUMENT}
%\rightline{Comment to {\tt d0$_-$run2$_-$qcd@fnal.gov,}}
%\rightline{{\tt d0-run2eb-038@fnal.gov}}
%\rightline{by Monday, December 15, 12pm}

\title{Measurement of the  $\bm{W+b}$-jet and $\bm{W+c}$-jet differential production cross sections in $\bm{p\bar{p}}$ collisions at $\bm{ \sqrt{s}=1.96}$~TeV}
\affiliation{LAFEX, Centro Brasileiro de Pesquisas F\'{i}sicas, Rio de Janeiro, Brazil}
\affiliation{Universidade do Estado do Rio de Janeiro, Rio de Janeiro, Brazil}
\affiliation{Universidade Federal do ABC, Santo Andr\'e, Brazil}
\affiliation{University of Science and Technology of China, Hefei, People's Republic of China}
\affiliation{Universidad de los Andes, Bogot\'a, Colombia}
\affiliation{Charles University, Faculty of Mathematics and Physics, Center for Particle Physics, Prague, Czech Republic}
\affiliation{Czech Technical University in Prague, Prague, Czech Republic}
\affiliation{Institute of Physics, Academy of Sciences of the Czech Republic, Prague, Czech Republic}
\affiliation{Universidad San Francisco de Quito, Quito, Ecuador}
\affiliation{LPC, Universit\'e Blaise Pascal, CNRS/IN2P3, Clermont, France}
\affiliation{LPSC, Universit\'e Joseph Fourier Grenoble 1, CNRS/IN2P3, Institut National Polytechnique de Grenoble, Grenoble, France}
\affiliation{CPPM, Aix-Marseille Universit\'e, CNRS/IN2P3, Marseille, France}
\affiliation{LAL, Universit\'e Paris-Sud, CNRS/IN2P3, Orsay, France}
\affiliation{LPNHE, Universit\'es Paris VI and VII, CNRS/IN2P3, Paris, France}
\affiliation{CEA, Irfu, SPP, Saclay, France}
\affiliation{IPHC, Universit\'e de Strasbourg, CNRS/IN2P3, Strasbourg, France}
\affiliation{IPNL, Universit\'e Lyon 1, CNRS/IN2P3, Villeurbanne, France and Universit\'e de Lyon, Lyon, France}
\affiliation{III. Physikalisches Institut A, RWTH Aachen University, Aachen, Germany}
\affiliation{Physikalisches Institut, Universit\"at Freiburg, Freiburg, Germany}
\affiliation{II. Physikalisches Institut, Georg-August-Universit\"at G\"ottingen, G\"ottingen, Germany}
\affiliation{Institut f\"ur Physik, Universit\"at Mainz, Mainz, Germany}
\affiliation{Ludwig-Maximilians-Universit\"at M\"unchen, M\"unchen, Germany}
\affiliation{Panjab University, Chandigarh, India}
\affiliation{Delhi University, Delhi, India}
\affiliation{Tata Institute of Fundamental Research, Mumbai, India}
\affiliation{University College Dublin, Dublin, Ireland}
\affiliation{Korea Detector Laboratory, Korea University, Seoul, Korea}
\affiliation{CINVESTAV, Mexico City, Mexico}
\affiliation{Nikhef, Science Park, Amsterdam, the Netherlands}
\affiliation{Radboud University Nijmegen, Nijmegen, the Netherlands}
\affiliation{Joint Institute for Nuclear Research, Dubna, Russia}
\affiliation{Institute for Theoretical and Experimental Physics, Moscow, Russia}
\affiliation{Moscow State University, Moscow, Russia}
\affiliation{Institute for High Energy Physics, Protvino, Russia}
\affiliation{Petersburg Nuclear Physics Institute, St. Petersburg, Russia}
\affiliation{Instituci\'{o} Catalana de Recerca i Estudis Avan\c{c}ats (ICREA) and Institut de F\'{i}sica d'Altes Energies (IFAE), Barcelona, Spain}
\affiliation{Uppsala University, Uppsala, Sweden}
\affiliation{Taras Shevchenko National University of Kyiv, Kiev, Ukraine}
\affiliation{Lancaster University, Lancaster LA1 4YB, United Kingdom}
\affiliation{Imperial College London, London SW7 2AZ, United Kingdom}
\affiliation{The University of Manchester, Manchester M13 9PL, United Kingdom}
\affiliation{University of Arizona, Tucson, Arizona 85721, USA}
\affiliation{University of California Riverside, Riverside, California 92521, USA}
\affiliation{Florida State University, Tallahassee, Florida 32306, USA}
\affiliation{Fermi National Accelerator Laboratory, Batavia, Illinois 60510, USA}
\affiliation{University of Illinois at Chicago, Chicago, Illinois 60607, USA}
\affiliation{Northern Illinois University, DeKalb, Illinois 60115, USA}
\affiliation{Northwestern University, Evanston, Illinois 60208, USA}
\affiliation{Indiana University, Bloomington, Indiana 47405, USA}
\affiliation{Purdue University Calumet, Hammond, Indiana 46323, USA}
\affiliation{University of Notre Dame, Notre Dame, Indiana 46556, USA}
\affiliation{Iowa State University, Ames, Iowa 50011, USA}
\affiliation{University of Kansas, Lawrence, Kansas 66045, USA}
\affiliation{Louisiana Tech University, Ruston, Louisiana 71272, USA}
\affiliation{Northeastern University, Boston, Massachusetts 02115, USA}
\affiliation{University of Michigan, Ann Arbor, Michigan 48109, USA}
\affiliation{Michigan State University, East Lansing, Michigan 48824, USA}
\affiliation{University of Mississippi, University, Mississippi 38677, USA}
\affiliation{University of Nebraska, Lincoln, Nebraska 68588, USA}
\affiliation{Rutgers University, Piscataway, New Jersey 08855, USA}
\affiliation{Princeton University, Princeton, New Jersey 08544, USA}
\affiliation{State University of New York, Buffalo, New York 14260, USA}
\affiliation{University of Rochester, Rochester, New York 14627, USA}
\affiliation{State University of New York, Stony Brook, New York 11794, USA}
\affiliation{Brookhaven National Laboratory, Upton, New York 11973, USA}
\affiliation{Langston University, Langston, Oklahoma 73050, USA}
\affiliation{University of Oklahoma, Norman, Oklahoma 73019, USA}
\affiliation{Oklahoma State University, Stillwater, Oklahoma 74078, USA}
\affiliation{Brown University, Providence, Rhode Island 02912, USA}
\affiliation{University of Texas, Arlington, Texas 76019, USA}
\affiliation{Southern Methodist University, Dallas, Texas 75275, USA}
\affiliation{Rice University, Houston, Texas 77005, USA}
\affiliation{University of Virginia, Charlottesville, Virginia 22904, USA}
\affiliation{University of Washington, Seattle, Washington 98195, USA}
\author{V.M.~Abazov} \affiliation{Joint Institute for Nuclear Research, Dubna, Russia}
\author{B.~Abbott} \affiliation{University of Oklahoma, Norman, Oklahoma 73019, USA}
\author{B.S.~Acharya} \affiliation{Tata Institute of Fundamental Research, Mumbai, India}
\author{M.~Adams} \affiliation{University of Illinois at Chicago, Chicago, Illinois 60607, USA}
\author{T.~Adams} \affiliation{Florida State University, Tallahassee, Florida 32306, USA}
\author{J.P.~Agnew} \affiliation{The University of Manchester, Manchester M13 9PL, United Kingdom}
\author{G.D.~Alexeev} \affiliation{Joint Institute for Nuclear Research, Dubna, Russia}
\author{G.~Alkhazov} \affiliation{Petersburg Nuclear Physics Institute, St. Petersburg, Russia}
\author{A.~Alton$^{a}$} \affiliation{University of Michigan, Ann Arbor, Michigan 48109, USA}
\author{A.~Askew} \affiliation{Florida State University, Tallahassee, Florida 32306, USA}
% ...
%
% visitor_addresses.tex                       18 October 2014
%  available symbols are:
%  $\ast, \dag, \ddag, \S, \P, $\|$, $\ast\ast$, \dag\dag, \ddag\ddag ,\#
%
\collaboration{The D0 Collaboration\footnote{with visitors from
%{alton}
$^{a}$Augustana College, Sioux Falls, SD, USA,
%{burdin}
$^{b}$The University of Liverpool, Liverpool, UK,
%{grohsjean,deterre}
$^{c}$DESY, Hamburg, Germany,
%{de la cruz-burelo}
$^{d}$Universidad Michoacana de San Nicolas de Hidalgo, Morelia, Mexico
%{partridge}
$^{e}$SLAC, Menlo Park, CA, USA,
%{hesketh}
$^{f}$University College London, London, UK,
%{luna-garcia}
$^{g}$Centro de Investigacion en Computacion - IPN, Mexico City, Mexico,
%{santos}
$^{h}$Universidade Estadual Paulista, S\~ao Paulo, Brazil,
%{meyer}
$^{i}$Karlsruher Institut f\"ur Technologie (KIT) - Steinbuch Centre for Computing (SCC),
D-76128 Karlsruhe, Germany,
%{patwa}
$^{j}$Office of Science, U.S. Department of Energy, Washington, D.C. 20585, USA,
%{cooke}
$^{k}$American Association for the Advancement of Science, Washington, D.C. 20005, USA,
%{borysova}
$^{l}$Kiev Institute for Nuclear Research, Kiev, Ukraine,
%{jabeen}
$^{m}$University of Maryland, College Park, Maryland 20742, USA
and
%{williams}
$^{n}$European Orgnaization for Nuclear Research (CERN), Geneva, Switzerland
%{montgomery}
%$^{?}$Thomas Jefferson National Accelerator Facility, Newport News, VA 23606, USA,
%{falkowski}
%$^{?}$Laboratoire de Physique Theorique, Orsay, FR,
%{hooper,kozminski}
%$^{?}$}Visitor from Lewis University, Romeoville, IL, USA.
%{weber}
%$^{?}$Universit{\"a}t Bern, Bern, Switzerland.
%{deceased}
%{zanabria}
%$^{?}$City Colleges of Chicago, Chicago, IL, USA}
%$^{\ddag}$Deceased.
}} \noaffiliation
\vskip 0.25cm

\date{February 4, 2015}

\begin{abstract}

We present a measurement of the cross sections for the associated production of a $W$ boson 
with at least one heavy quark jet, $b$ or $c$, in proton-antiproton collisions.
% The  measurements are done using 
Data corresponding to an integrated luminosity of 8.7~fb$^{-1}$ recorded with the D0 detector 
at the Fermilab Tevatron \ppbar~Collider at $\sqrt{s}=1.96$~TeV are used to measure the cross sections differentially 
% The measurements are presented  differentially 
 as a function of the jet transverse momenta in the range 20 to 150 GeV. These results are compared 
% to  perturbative QCD theory calculations as well
 to calculations of perturbative QCD theory as well
 as predictions from Monte Carlo generators.

\end{abstract}

\maketitle

%\linenumbers

Measurement of the production cross section of a $W$ boson in association with a $b$ or $c$-quark jet provides a stringent test of quantum chromodynamics (QCD).
At hadron colliders, the associated production of a heavy quark with a  $W$ boson can also be a significant background to  
%more
 rare standard model (SM) processes,
for example, production of top quark pairs~\cite{toppair}, a single top quark~\cite{s-top}, and a $W$ boson in association  with a Higgs boson  decaying to 
two $b$ quarks~\cite{HW}, as well as for new physics processes, e.g., supersymmetric scalar top quark production~\cite{stop_susy}. %decaying via $ \tilde{t} \to c\tilde{\chi}$. 

%Additionally, the cross section of  $p \bar{p} \to W + c- jets $ is directly sensitive to the strange quark PDF
% and can be used for the s-quark PDF extraction.
 The dominant processes contributing to \wc production are $q g \to W c $ %(with $q = d,s,b$)
  and $ q \bar{q}' \to W g$ followed by $g \to c \bar{c}$.
%  (with $q = d,s,b$ and $ \bar{q}' = \bar{u}, \bar{c}$).
%
  The production cross section for the first process is sensitive to the quark and gluon  parton density functions (PDFs).
%  and to CKM matrix element, which  are  now known quite well: $ |V_{cd}|^2 \approx 0.04 ,  |V_{cb}|^2 \approx 0.002 , |V_{cs}|^2 \approx 0.95$ \cite{PDG}. 
  Since the $c$-$b$  quark Cabibbo-Kobayashi-Maskawa
%  mixing angle
matrix element 
 is very small ($|V_{cb}|^2 \approx 0.0016$)~\cite{PDG},
  the contribution of a $b$-quark initial state in the PDF  is negligible. 
Comparing the $d$-quark and $s$-quark PDFs, the probability of interaction of a gluon with an $d$-quark is greater than that with a $s$-quark, 
% The $d$-quark PDF is higher than 
%   that of the $s$-quark, 
but the CKM matrix element suppresses $d \to c$ transitions since $|V_{cd}|^2 \approx 0.04$.
  As a result, the expected contributions from $s$-quark and $d$-quark initial states for a jet transverse momentum $p_T^{\rm jet}>20$~GeV  at the Tevatron
  are around $85\%$ and $15\%$, respectively~\cite{D0_2008}. 
  According to the {\sc alpgen+pythia}~\cite{alpgen,pythia} simulation for \wc events, the  contribution from $qg\to W+c$
%  dominates the
% %  whole 
% % jet%
% entire  
% $p_T$ region $>20$ GeV and $<100$ GeV
dominates the entire $20 < p^{\rm jet}_T < 100$ GeV region
%   covered by this measurement
 with the contribution from $q \bar{q}'\to W+c\bar{c}$ events increasing from about $25\%$ to $45\%$
%   at jet $p_T$ ranges from 20 to 100 GeV.
 as jet $p_T$ increases from 20 to 100 GeV.
  Measurement of the $ p \bar{p} \to W + c$-jet  differential cross section should provide 
  information about the $s$-quark PDF.
%  at the energy scale $Q>20$ GeV.
  This PDF has been measured directly only in fixed target neutrino-nucleon deep inelastic scattering experiments 
%using
 at  relatively 
  low momentum transfer $Q\lesssim 15-20$ GeV~\cite{NuTev_1,NuTev_2,CCFR_1,CCFR_2,CharmII,CDHS}.
  %
  %This is additional to an indirect source of $s$-quark PDF information coming from neutral and charged currents in deep-inelastic scattering \cite{HERA_s},
  %and  a significant extension of CCFR/NuTeV neutrino nucleon scattering experiments ~\cite{neutrino}, which
  %mainly cover region up to $Q\lesssim 15-20$ GeV.
  A probe of the $s$-quark PDF at the Tevatron tests the universality of $s(x,Q^2)$, where $x$ is the  Feynman variable~\cite{Feynman69}, and its QCD evolution up to $Q^2\simeq 10^4$ GeV$^2$.

There are only  a few previous measurements of the \wc 
cross section at hadron colliders, performed by the D0~\cite{D0_2008}, CDF~\cite{CDF_2008, CDF_2013}, ATLAS~\cite{atlas_strange}, and CMS~\cite{CMS_2014}
Collaborations.
 The previous D0 and CDF measurements are inclusive; the CMS  and ATLAS inclusive results were augmented by distributions in the pseudorapidity of the lepton from $W$ decay. 
% All measurements are in agreement with the 
% perturbative next-to-leading order (NLO) QCD predictions ~\cite{assoc_wb1, assoc_wb}
% within typical theoretical uncertainties of 15--30\%.
 It is important to note that the measurements listed above are performed by
% exploiting
 requiring opposite electric charges of a 
% soft lepton inside a jet and a lepton from $W$ decay
soft lepton inside a jet from semileptonic
% $c$-quark
charmed hadron
 decay with a lepton from $W$ decay elsewhere in the event, and measuring the cross section for ``opposite-sign''(OS) minus ``same-sign''(SS) events.
% A requirement of opposite signs for the leptons and subtraction of  events with the same signs removes 
% %all of  the
%  sign-symmetric backgrounds,
% but it also excludes the sign-symmetric $W+c\bar{c}$ events that become significant at high jet $p_T$, 
% i.e. in the region important for  supersymmetry searches and the Higgs boson studies. 
% This method also has small data statistics due to a low lepton tagging efficiency of the $b/c$-quark jet.
A requirement of opposite signs for the leptons and subtraction of events with the same signs suppresses the sign-symmetric backgrounds as well as $W+c \bar c$ events due to gluon splitting, which become significant at high jet $p_T$.
% The NLO predictions included contributions from gluon splitting.
 All measurements
are in agreement with the perturbative next-to-leading
order (NLO) QCD predictions~\cite{assoc_wb1, assoc_wb} that include contributions from gluon splitting within total 
theoretical uncertainties of 15--30\%.
% This measurement does not require a soft lepton.
% The present measurement does not require a soft lepton and thus does not suppress the contribution from the gluon spliting process.

Measurements of the inclusive \wb  cross-sections have been reported by the CDF ~\cite{CDF_meas}, D0~\cite{D0_meas}, and ATLAS~\cite{Atlas_meas} Collaborations.
%The CDF measurement of $ 2.74 \pm 0.27 (stat) \pm 0.42 (syst)$ pb(The result is found to be larger than the NLO predictions $1.22\pm 0.14 (syst)$ pb, 
%and the ATLAS measurement of $10.2\pm 1.9 (stat)\pm 2.6 (syst) $pb( Fig.~\ref{fig:atlas_result}) are both found to be larger than the corresponding 
% theoretical cross-sections calculated, but in agreement within large(30-40\%) theoretical uncertainties. 
% The CDF result is noticeably higher
The CDF result is approximately $3\sigma$ higher
 than the NLO predictions while the D0 and ATLAS measurements agree with the theory within large (30--40\%) theoretical uncertainties.
A dominant ($\simeq 85\%$) contribution to $W+b$-jet production at the Tevatron is due to the $q \bar{q}' \to W+g~(g\rightarrow b \bar{b})$ process while
the remaining contribution 
% comes from 
arises from
the  $b\bar{q} \to Wb\bar{q}'$ process \cite{assoc_wb}, with a negligible contribution from single top quark production.

%In this letter,  
We present, for the first time, measurements of 
%fully inclusive
 \wc and \wb 
%cross sections  done
%made  differentially in jet $p_T$ and which
differential cross sections as a function of  jet $p_T$, where no requirement of a soft lepton within a jet is made, and that are therefore sensitive to the gluon splitting contributions.
% that  are sensitive to the gluon splitting contributions.
The $W$ boson candidates are identified in the $\mu+\nu$ decay channel. % whereas a small fraction of selected events arises from leptonical decaying tau leptons.

The data used in this analysis were collected between July 2006 and September 2011 using the D0 detector at the Fermilab
Tevatron Collider at $\sqrt{s}~=~1.96$~TeV, and correspond to an integrated luminosity of 8.7~\fb. 
% We first briefly describe the main components of the D0 Run II detector
% \cite{run2det} relevant to this analysis.
 The D0 detector~\cite{run2det} has a central tracking system consisting of a 
silicon microstrip tracker (SMT)~\cite{layer0} and a central fiber tracker, 
both located within a 1.9~T superconducting solenoidal 
magnet, which are optimized  for tracking and 
vertexing at pseudorapidities $|\eta|<3$ and $|\eta|<2.5$, respectively~\cite{coord}.
A liquid argon and uranium calorimeter has a 
central section (CC) covering pseudorapidities $|\eta| \lesssim 1.1$, and two end calorimeters (EC) that extend coverage 
to $|\eta|\approx 4.2$, with all three housed in separate 
cryostats~\cite{calopaper}. An outer muon system covering  $|\eta|<2$ 
consists of a layer of tracking detectors and scintillation trigger 
counters in front of 1.8~T iron toroids, followed by two similar layers 
after the toroids. 
Luminosity is measured using plastic scintillator 
arrays located in front of the EC cryostats. 
%The trigger and data  acquisition systems are designed to accommodate the high instantaneous luminosities.
% of Run II of the Tevatron. 

The $W+b/c$ candidate events are chosen by selecting single muon or muon+jet signatures with a three-level trigger system. 
The trigger efficiency has been estimated using $Z\to \mu^{+}\mu^{-}$(+jets) events in data. 
The trigger efficiency is parametrized as a function of muon $p_T$ and $\eta$ and is on average $\approx 70\%$.
%  for this data sample.
%The simulation is corrected for the trigger efficiencies measured in data as described in Ref.~\cite{wh_prd}.
%The trigger efficiency has been estimated using $Z\to \mu\mu$(+jets) events in data.

Offline event selection requires a reconstructed  $p\bar{p}$~interaction primary vertex (PV) 
that has at least three associated tracks and is located within $60$~cm of the center of the detector along the beam direction.  
The vertex selection for $W+b/c$ events is approximately $99\%$ efficient as measured in simulation.
% \CHECK{$99\%$ efficient} as measured in simulations.

%The selection criteria
We require a muon candidate to be reconstructed from hits in the muon system and matched to a reconstructed track in the
central tracker~\cite{muon_det}. The transverse momentum of the muon must satisfy $p_T^{\mu}>20$~GeV, with  $|\eta^{\mu}|<1.7$. 
Muons are required to be spatially isolated from other energetic particles using information from the central tracking detectors and calorimeter~\cite{wh_prd}. Muons from cosmic rays are rejected by applying a timing criterion on the hits in the scintillator layers of the muon system 
and by applying restrictions on the displacement of the muon track with respect to the PV. The muon reconstruction efficiency is  $\approx90\%$.

Candidate $W+\textrm{jets}$ events are selected by requiring at least one reconstructed jet with pseudorapidity $|\eta^{\rm jet}| < 1.5$ 
and $p_T^{\rm jet} > 20$~GeV. Jets are reconstructed from energy deposits in the calorimeter using the iterative midpoint cone algorithm~\cite{jet_algo} 
with a cone of radius $R=\sqrt{\Delta y^2+\Delta\phi^2}=0.5$
%  in $y$-$\phi$ space
~\cite{coord}. The energies of jets are corrected for detector response, the presence of noise 
and multiple \ppbar~interactions~\cite{jes_nim}.
%, and for energy deposited outside of the jet reconstruction cone \cite{jes_nim}.
 To enrich the sample with $W$ bosons, events are required to have missing transverse energy 
\cite{jes_nim} \etmiss $>25$~GeV due to the neutrino escaping detection. We
require that the $W$ boson candidates have a transverse mass $M_T>40$~GeV~\cite{mt}.

%Background processes for this analysis % are productions of
%include
%$W+\textrm{light jets}$, $Z/\gamma^*+\textrm{jets}$, $t{\bar t}$ and single top quark  events, 
%diboson $VV$ ($V=W,Z$) events, and multijet events with a jet misidentified as a muon. 
Backgrounds for this analysis include events from the production of $W+\textrm{light parton jets}$, $Z/\gamma^*+\textrm{jets}$, $t{\bar t}$,
 single top quark, diboson $VV$ ($V=W,Z$) and QCD multijets in which a jet is 
misidentified as a muon.
The $W+c$ and $W+b$ signal and the background processes excluding multijet are simulated using a combination of  
\textsc{alpgen}~\cite{alpgen} and \textsc{pythia}~\cite{pythia} MC event generators with \textsc{pythia} providing parton showering and hadronization. 
We use \textsc{pythia}  with CTEQ6L1~\cite{pdf_cteq6M} PDFs.
%and perform a detailed \textsc{geant}-based~\cite{geant} simulation of the D0 detector. 
{\sc alpgen} generates multi-parton final states using tree-level matrix elements (ME). When interfaced with
{\sc pythia}, it employs the MLM scheme \cite{alpgen} to treat ME partons produced from showering in {\sc pythia}.
For the signal process, we also use the {\sc sherpa} MC generator \cite{sherpa} that matches  partons from the leading-order ME with up to two real parton emissions
to the parton-shower jets according to the CKWK matching scheme~\cite{ckkm}.
The generated events are processed through a {\sc geant}-based~\cite{geant} simulation of the
D0 detector geometry and response.
To accurately model the effects of multiple $p\bar{p}$ interactions
and detector noise, events from random $p\bar{p}$
crossings with a similar instantaneous luminosity spectrum as
in data are overlaid on the MC
events. These MC events are then processed using the
same reconstruction code as for the data.
The MC events are also weighted
to take into account the trigger efficiency 
% in data,
 and small observed differences between MC and data in the distributions of 
the instantaneous luminosity and of the $z$ coordinate of the $p\bar{p}$ collision vertex.

The $V{\rm +jets}$  processes are normalized to the total inclusive $W$ and $Z$-boson cross sections calculated at NNLO (next-to-next-to-leading order)~\cite{hamberg}.
The $Z$-boson $p_T$ distribution is modeled to match the distribution observed in data~\cite{zpt_xsec}, taking into account the dependence on the number of reconstructed jets.
 To reproduce the $W$-boson $p_T$ distribution in simulated events, we use the product of the measured $Z$-boson $p_T$ spectrum 
%and
times the ratio of $W$ to $Z$-boson $p_T$\ distributions at NLO
%is used as a correction factor
~\cite{zpt_xsec, wpt_incl}.  
The NLO+NNLL (next-to-next-to-leading log) calculations are used to normalize \ttbar\ production~\cite{mochuwer}, 
while single top quark production is normalized to NNLO predictions~\cite{single_top}. 
The NLO $WW$, $WZ$, and $ZZ$ production cross section values are obtained with the \textsc{mcfm} program~\cite{mcfm}. 
%For the $W$+heavy-flavor jet ($b$ or $c$ quark) events, the ratio of the \textsc{alpgen} prediction to the \textsc{NLO} prediction for $W+b\bar{b}$ and ~$W+c\bar{c}$ is obtained from \textsc{mcfm}~\cite{mcfm} and applied as a correction factor.  
%The simulation is also corrected for the trigger efficiencies measured in data.
The multijet background contribution is estimated from data using the ``matrix method''~\cite{wh_prd}. 
% In the $W$ boson muon decay channel, 
For the final states studied here 
the multijet background is small ($\lesssim2\%$) and arises mainly from the semileptonic decays of heavy quarks in which the muon 
satisfies the isolation requirements. 
%We require that the $W$ boson candidates have a transverse mass $M_T$~\cite{mtw} satisfying \mbox{$40~\text{GeV}+~\frac{1}{2}\etmiss < M_T < 120$~GeV} to suppress multijet background and mis-reconstructed events. The average efficiency determined in simulation for a $W+b$ signal to pass these requirements is about $82\%$. 
To reduce the contribution from $t{\bar t}$ production that increases with jet $p_T$, we restrict the scalar sum of all the jet $p_T$ values 
($H_T$) to be less than 175 GeV. 
This requirement reduces the $t{\bar t}$ fraction by a factor $1.5-2$, depending on  jet $p_T$, and
has signal efficiency greater than  95\% except in the highest PT bin where it falls to 82\%.
% above $82\%$.
 The $t \bar t$ fraction after the $H_T$ cut varies between 5 and 20\% with increasing  jet $p_T$.
% \CHECK{This cut reduces the $t{\bar t}$ fraction by a factor $1.5-2$ (depending on the jet $p_T$ bin) with the signal efficiency of more $82\%$.}
%We vary this cut by $\pm15\%$ to estimate a systematic uncertainty.

Identification of $b$ and $c$ jets is crucial for this measurement.
Once the inclusive $W+{\rm jets}$ sample is selected, at least one jet is required to be taggable, i.e.
it must contain at least   two tracks each with at least one hit in the SMT, $p_T>1$~GeV for the highest-$p_T$ track 
and $p_T>0.5$~GeV for the next-to-highest $p_T$ track.
%  The efficiency for a jet to be taggable is about $90\%$ in the selected phase space.
%The jet is required to have at least two associated tracks with $p_T>0.5$~GeV
%with at least one hit in the SMT.
%The track with the highest $p_T$ must have $p_T>1.0$~GeV.
These criteria ensure
%  that there is 
 sufficient information to
classify the jet as a heavy-flavor candidate and have a typical efficiency
of about 90\%.  Light parton jets (those resulting from light quarks or gluons) are suppressed
using a dedicated artificial neural network ($b$-NN)~\cite{bid_nim}
that exploits the longer lifetimes of heavy-flavor hadrons relative to
their lighter counterparts.
The inputs to the $b$-NN include several characteristic 
% qualities 
 quantities of the
jet and associated tracks to provide a continuous output
value that tends towards one for $b$ jets and zero for the light
jets. The $b$-NN input variables providing most of the discrimination are
the number of
reconstructed secondary vertices (SV) in the jet, the invariant mass of charged
particle tracks associated with the SV ($M_{\rm SV}$),
the number of tracks used to reconstruct the SV, the two-dimensional
decay length significance of the SV in the plane transverse
to the beam, a weighted combination of the tracks' transverse impact parameter significances,
and the probability that the tracks associated with the jet originate from the $p\bar{p}$ interaction vertex,
which is referred to as the jet lifetime probability (JLIP).
The jet is required to have a
$b$-NN output greater than  $0.5$.
% Depending on jet $p_T$,
For jet $p_T$ in a range between 20 and 150 GeV this selection is ($36-47$)\% efficient for $b$-jets and ($8-11$)\% efficient for $c$ jets
with
% fractional
relative systematic uncertainties of ($4.2-6.5$)\% for both the $b$ jets 
and  $c$ jets. 
The systematic uncertainty is obtained from a comparison of the heavy flavor tagging efficiencies in data and MC as described in \cite{bid_nim}.
Only $0.2-0.4$\% of light jets are misidentified as heavy-flavor jets and comprise
$7$\% to $15$\% of the final sample, with a larger fraction at lower jet $p_T$.
%  at small jet $p_T$.
% \CHECK{
% Depending on jet $p_T$, this selection is ($40-52$)\% efficient for $b$-jets and ($9-12$)\% efficient for $c$-jets
% with systematic uncertainties of ($x1-x2$)\% for the $b$-jets 
% and ($y1-y2$)\% for the $c$-jets, respectively.
% The systematic uncertainty is obtained from a comparison of the heavy flavour tagging efficiencies in data and MC as described in \cite{bid_nim}.
% Only $0.2-0.4$\% of light jets are misidentified as heavy-flavor jets, comprising
% $7$\% to $20$\% of the final sample, with a larger fraction at small jet $p_T$.
% }
%
In addition to the $b$-NN output, we obtain further information by combining the $M_{\text{SV}}$ and JLIP variables,
which  provide good discrimination between $b$, $c$, and light quark jets due to their different masses~\cite{Zb, gamma_b}.
% The two variables together take into account the kinematics of the event and, 
%In order to further improve the separation power, they are
%combined in a single variable
We form a single variable discriminant  ${\cal D}_{\text{MJL}}~=\frac{1}{2}~\left(M_{\text{SV}}/(5~\text{GeV}) - \ln(\text{JLIP})/20 \right )$~\cite{Zb} 
%A  criterion
%  for an event to pass at least loose
and require ${\cal D}_{\text{MJL}}>0.1$  to remove poorly reconstructed events  and reduce the number of  light-jet events. 
The efficiency for signal events to pass this selection is  $98\%$ for $b$-jets and $97\%$ for $c$-jets.
% \CHECK{The efficiency for signal events to pass this selection is about $98\%$ for $b$-jets and $97\%$ for $c$-jets.}

\begin{figure}[!h]
\includegraphics[width=0.5\textwidth]{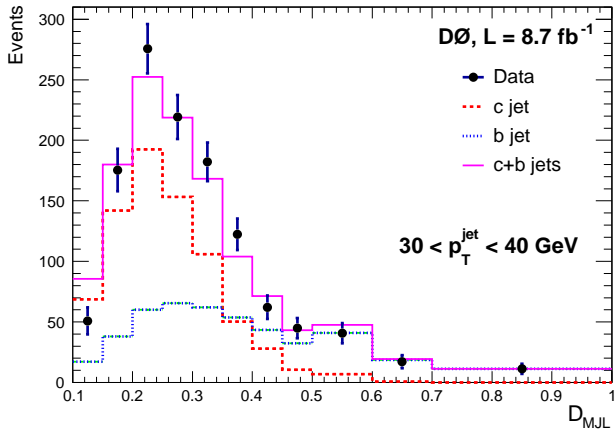}
\caption{(Color online) Distribution of the $D_{\rm MJL}$ discriminant
  after all selection criteria (including $b$-NN output $>0.5$) for a representative bin of $30<p_{T}^{\rm jet}< 40$~GeV.
  The contributions from background events are subtracted from data before the fit.
%  The expected contribution from the light-jet component has been subtracted from the data.
  The distributions for the $c$-jet and $b$-jet templates (with statistical uncertainties)
  are shown normalized to their respective fitted fractions.}
 \label{fig:fit1}
\end{figure}

After all selection requirements, 5260 events remain in the data sample.
%\CHECK{XXX} events remain in the data sample.
We measure the fraction of $W+c$ and $W+b$ events in the selected sample by performing a binned maximum likelihood fit 
to the observed data distribution of the ${\cal D}_{\text{MJL}}$ discriminant in bins of jet $p_T$, as shown in Fig.~\ref{fig:fit1} for the bin $30 < p_T^{\rm jet} < 40$~GeV.
The templates for $W+b$ and $W+c$ jets are taken from the simulation. 
Expected contributions from the background processes are subtracted
%Expected contributions from $W+$light jets, $Z$+jets, single top quark, $t\bar{t}$, diboson, and multijet production are subtracted 
 from the ${\cal D}_{\text{MJL}}$ distribution in data before the fit.
% The relative contribution from $W+$light jets (“light/(light$+b+c$)”)
The ratio of the  $W+$light parton jets to $W+$ all jet flavors
 has been estimated using {\sc alpgen+pythia} MC events 
taking into account the data-to-MC correction factors as described in Ref.~\cite{bid_nim},
and has been cross checked in data using looser cuts on $b$-NN output in the range from 0.15--0.3.

\begin{figure}[!h]
\includegraphics[width=0.45\textwidth]{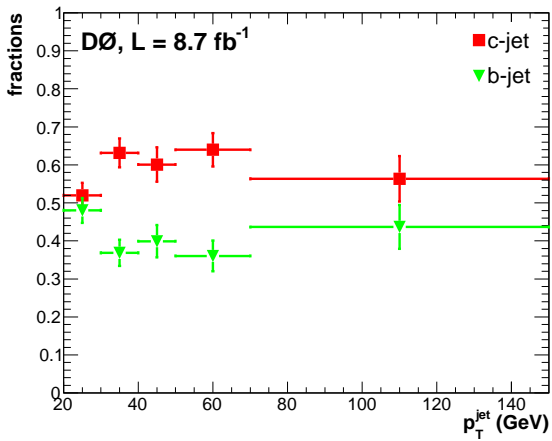}
\caption{(Color online)
The $b$- and $c$-jet fractions (their total sum is normalized to 1.0) versus jet $p_T$ with total uncertainty from the $D_{\rm MJL}$ fit.}
 \label{fig:pt_fraction_0}
\end{figure}

The fractions of $W+b$ and $W+c$ events after subtraction of 
%$W+$ light-jets and
 background contribution are shown in Fig.~\ref{fig:pt_fraction_0} as a function of jet $p_T$.
The  relative uncertainties on the fractions obtained from the fit range within ($7-13$)\% for $W+b$ and ($6-11$)\% for $W+c$.
This includes the uncertainty due to $W+b$ and $W+c$ template shapes, studied in a previous analysis~\cite{gamma_c}.
The contributions from the background events are varied within uncertainties on their predicted cross sections,
and these uncertainties are propagated into the extracted signal fractions.
The uncertainty due to the light parton jets template shape
% has been studied previously
 is taken from Ref.~\cite{gamma_b}.
The overall relative uncertainties on the subtracted backgrounds range within ($4-6$)\% for $W+b$ and ($3-4$)\% for $W+c$.

%The measured cross sections are presented at the particle level by correcting for detector acceptance, selection-efficiencies, and $b$-jet identification. 
%Trigger and object reconstruction efficiencies are parametrized as function of their kinematics $\eta$~\ref{wh_prd}, 
%migration effects of the restricted phase space and efficiency of selection requirements are corrected.
%
%

We  apply corrections to the measured number of signal events
to account for the detector and kinematic acceptances and selection efficiencies using simulated samples
of $W+b(c)$-jet events.
In these calculations, we apply the following selections at the particle level:
at least one $b(c)$-jet with  $p_T^{\text{$b(c)$-jet}}>20$~GeV, $|\eta^{\text{$b(c)$-jet}}|<1.5$,
a muon with $p_T^\mu~>~20$~GeV and $|\eta^\mu|<1.7$, and a neutrino with $p_T^\nu>25$~GeV. 
% Here, the particle level includes all stable particles as defined in Ref.~\cite{particle}.
In the following, we quote our cross section results for this restricted phase space as a fiducial cross section.

%
%%% c & b fraction fit
%

%
%%% c & b fractions vs jet pT
%
% \begin{figure}[!h]
% \includegraphics[width=0.45\textwidth]{Fractions_1.eps}
% % \includegraphics[width=0.45\textwidth]{Fractions.eps}
% \caption{(Color online)
% The $b$- and $c$-jet fractions (their total sum is normalized to 1.0) versus jet $p_T$ with total uncertainty from the $D_{\rm MJL}$ fit.}
%  \label{fig:pt_fraction_0}
% \end{figure}

%Signal events are generated using the {\sc pythia+alpgen} event generators, and processed through 
%a {\sc geant}-based~\cite{Geant} simulation and events reconstruction as described above.
The acceptance is defined by the selection requirements in jet and muon transverse momenta and pseudorapidities.
% The acceptance is driven by the detector geometry, kinematic selections
% in jet and muon transverse momenta and rapidities, and bin-to-bin
% migration effects due to
%(applied to avoid edge effects in the calorimeter regions used for the measurement)
%and $\phi_{\rm det}$ in the central rapidity region
%(to avoid periodic calorimeter module boundaries \cite{d0det}
%that bias the EM cluster energy and position measurements),
% jet energy corrections, and bin-to-bin migration effects due to the jet and muon finite energy and angular resolutions.
Correction factors to account for small differences
between jet-$p_T$ and rapidity spectra in data and simulation are estimated, and used as weights to
create a data-like MC sample. The differences between acceptance corrections obtained with standard and 
corrected MC samples are taken as a systematic uncertainty of up to 3\% at low jet $p_T$. 
An additional systematic uncertainty of up to 4\% is due to uncertainties in the jet energy correction and resolution.
For 
%jet $p_T$ in the range between 
 $20 < p_T^{\rm jet} < 150$  GeV the product of acceptance and muon selection efficiency varies within ($50-65$)\% with a relative systematic uncertainty of ($3 - 5$)\%.
%The total selection efficiency of the muon selection criteria is \CHECK{about $80\%$}, parametrized as a function of the detector azimuthal angle and pseudorapidity.
The systematic uncertainty on the muon selection efficiency is about 2\% and is obtained from a comparison of the muon efficiencies 
in $Z\to \mu\mu$ events in data and MC.
Uncertainties on $b$-jet identification are determined in simulations and data by 
using $b$-jet-enriched samples \cite{bid_nim} and are about (2 -- 5)\%  per jet. 
%The uncertainties due to lepton identification are about $2\%$. 
The integrated luminosity is known to a precision of $6.1\%$~\cite{lumi}. 
%The uncertainty of the template fit is estimated by varying the normalization 
%and shape from the data corrections of the $W$ boson processes and the fit parameters (about $6\%$). 
By summing the uncertainties in quadrature we obtain a final total systematic uncertainty on the cross section measurements of ($11-18$)\%
depending on jet $p_T$ and final state.

\begin{figure}[!h]
\includegraphics[width=0.5\textwidth]{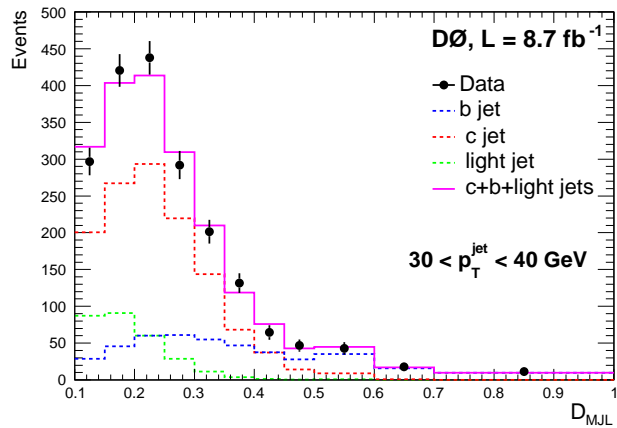}
\caption{(Color online) 
%Distribution of $D_{\rm MJL}$ discriminant
%  after all selection criteria (including $b$-NN output $>0.3$) for a representative bin of $30<p_{T}^{\rm jet}< 40$~GeV.
%  Contribution from background events is subtracted from data before the fit.
%  The distributions for the light jet, $c$-jet and $b$-jet templates (with statistical uncertainties)
%  are shown normalized to their respective fitted fractions. 
  Similar to Fig.~\ref{fig:fit1} but for events selected with $b$-NN output $>0.3$. This alternative selection  is used as a cross-check of the main results.}
 \label{fig:fit2}
\end{figure}

To check the stability of the results, the \wb and \wc cross sections have been remeasured using a looser $b$-NN selection,
$b$-NN output $>0.3$, and with the light parton jet  fraction included as an additional fit parameter, 
thus increasing the data statistics and the background fractions.
The fractions of $b$- and $c$-jets are obtained from the maximum likelihood fit of 
the light, $b$- and $c$-jet templates to ${\cal D}_{\text{MJL}}$ distribution in data  as shown in  Fig.~\ref{fig:fit2}.
%
% Similarly, 
We also vary the default $H_T$ cut by $\pm15$~GeV and remeasure the cross sections.
In both  cross-checks, the default and new cross sections are found to be in agreement within  uncertainties that include the correlation between the
two measurements.
%
%%% c & b fraction fit
%

\begin{figure}[!h]
\vskip -3mm
\includegraphics[width=0.45\textwidth]{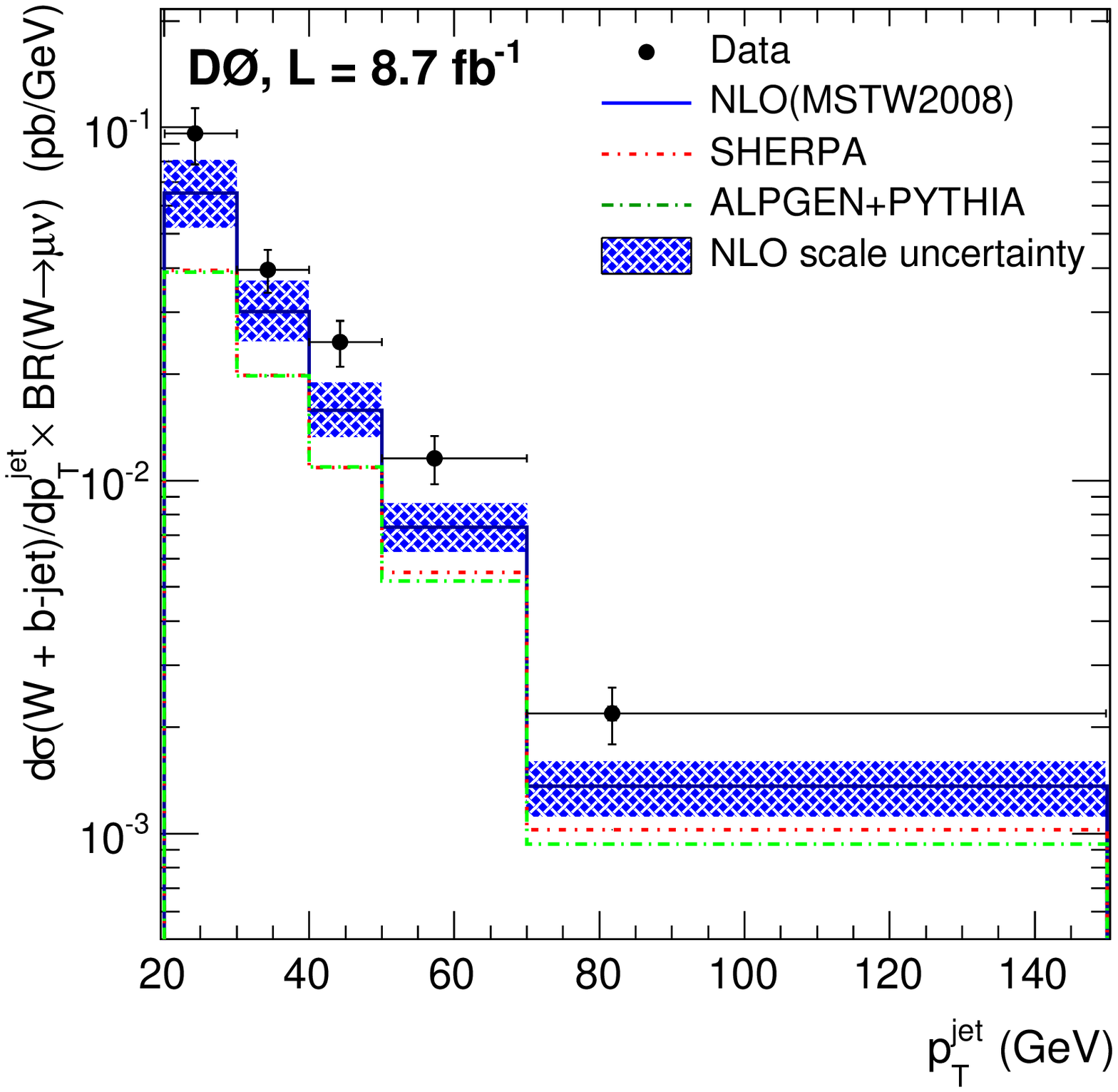}
\caption{(Color online)
The \wb differential production cross section times $W \to \mu\nu$ branching fraction  as a function of jet $p_T$. 
The uncertainties on the data points include statistical and systematic contributions added in quadrature.
The measurements are compared to the NLO QCD calculations~\cite{mcfm} using the MSTW2008 PDF set \cite{mstw}
(solid line). The predictions from {\sc sherpa}~\cite{sherpa} and {\sc alpgen}~\cite{alpgen} 
are shown by the dotted and dashed lines, respectively.
%This includes the theoretical scale uncertainties. The uncertainties on the points in
%data include both statistical (inner line) and the full uncertainties (the entire line) 
}
 \label{fig:xsce_log_wb}
\end{figure}

\begin{figure}[!h]
\includegraphics[width=0.45\textwidth]{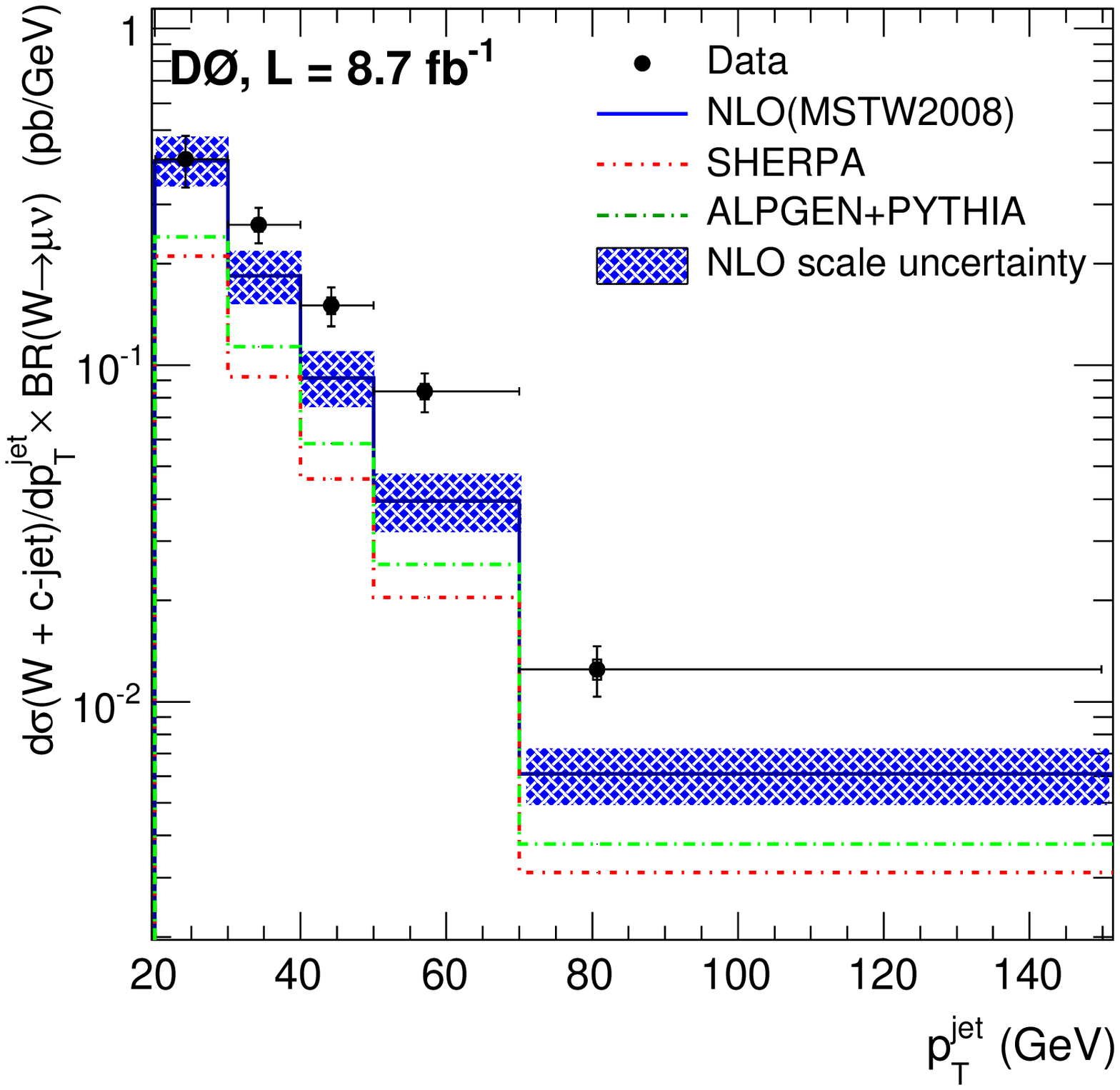}
\caption{(Color online)
The \wc differential production cross section times $W \to \mu\nu$ branching fraction as a function of jet $p_T$. 
The uncertainties on the data points include statistical and systematic contributions added in quadrature.
The measurements are compared to the NLO QCD calculations~\cite{mcfm} using the MSTW2008 PDF set \cite{mstw}
(solid line). The predictions from {\sc sherpa}~\cite{sherpa} and {\sc alpgen}~\cite{pythia} 
are shown by the dotted and dashed lines, respectively.
}
\label{fig:xsce_log_wc}
\end{figure}

In Figs.~\ref{fig:xsce_log_wb} and \ref{fig:xsce_log_wc} and 
Tables \ref{tab:res_b} and \ref{tab:res_c},
we present the \wb and \wc differential production cross sections times $W\to \mu\nu$ branching fraction 
for the fiducial phase space defined 
by $p_T^{\mu}>20$~GeV, $|\eta^\mu|<1.7$, $p_T^{\nu}>25$~GeV, and with at least one $b(c)$-jet with $p_T^{\rm jet}>20$~GeV and $|\eta^{\rm jet}|<1.5$.
The cross sections are presented differentially in five $p_T^{\rm jet}$  bins in the region $20-150$ GeV.
The data points are plotted at the value of $p_{T}^{jet}$ for which the value of a smooth
function describing the cross section equals the averaged
cross section in the bin~\cite{laferty}.
The cross sections are compared to predictions from NLO QCD~\cite{mcfm}  and two MC generators,  {\sc sherpa} and {\sc alpgen}.
The NLO predictions are made using the MSTW2008 ~\cite{mstw} and CT10~\cite{cteq} PDF sets.
We calculate the NLO QCD prediction using \textsc{mcfm} with central values of renormalization and fragmentation scales $\mu_{r}=\mu_{f}=M_W$
and with the $b$-quark and $c$-quark masses $m_b = 4.75$~GeV  and $m_c = 1.5$~GeV, respectively. 
Uncertainties are estimated by varying $\mu_{r}$ and $\mu_{f}$ independently  by a factor of two in each direction.  
%The \textsc{mcfm} calculation predicts $\sigma(W+b)\cdot  {\cal B}(W \to \mu \nu)= 1.34~^{+0.40}_{-0.33}~(\textrm{scale}) \pm 0.06~(\textrm{PDF})~^{+0.09}_{-0.05}~(m_b)$~pb.
%Predictions obtained using \textsc{sherpa} v1.4 and CTEQ6.6 PDFs~\cite{pdf_cteq6M} lead to a value $1.21 \pm 0.03 \thinspace(\textrm{stat.})$~pb. 
%Using \textsc{madgraph5}~\cite{mg5} with CTEQ6L1 PDFs, we obtain $1.52 \pm 0.02 \thinspace(\textrm{stat.})$~pb. 
%Uncertainties for scale variations, PDFs, and the $b$-quark mass are on the order of about 30\%.

%
%%% WHF cross sections in data
%

%The $W+b(c)$ production cross sections times $W\to \mu\nu$ branching fraction is calculated by dividing the number of signal events measured 
%by integrated luminosity ($\mathcal{L}$), acceptance ($\mathcal{A}$), and efficiencies ($\epsilon$) of the selection requirements:
%\begin{equation}
%  \sigma(W+b) \cdot  {\cal B}(W \to \ell \nu)  = \frac{N_{W+b} }{{\cal L} \cdot {\cal A} \cdot  \epsilon},
%\label{eq:qcd_xsec}
%\end{equation}
%where $\epsilon$ is given by the product of the trigger, object reconstruction, and selection efficiencies.

%We first present results separately for the muon channel and electron channel because they are performed in slightly different requirements 
%on the phase space of the lepton and then combine using a common phase space. 

% The data points are plotted at the value of for which the value of the smooth
%function describing the cross section equals the averaged
%cross section in the bin [24].
%

\begin{table*}[!]
\begin{center}
\caption{The \wb production cross sections times $W \to \mu\nu$ branching fraction, ${\rm d}\sigma/{\rm d}p_T^{\rm jet}$, 
together with statistical uncertainties ($\delta_{\text{stat}}$) and  total systematic uncertainties ($\delta_{\text{syst}}$). 
The column $\delta_{\rm tot}$ shows total experimental uncertainty obtained by adding $\delta_{\text{stat}}$
and $\delta_{\text{syst}}$ in quadrature.
The last three columns show theoretical predictions obtained using NLO QCD with MSTW PDF set, and two MC event generators,
{\sc sherpa} and {\sc alpgen}.
}
\label{tab:res_b}
\begin{tabular}{ccccccccc} \hline \hline
~$p_T^{\rm jet}$ bin~ & ~$\la p_T^{\rm jet}\ra$~ & \multicolumn{7}{c}{${\rm d}\sigma/{\rm d}p_T^{\rm jet}$ (pb/GeV) } \\\cline{3-9}
 (GeV) & (GeV) & Data & $\delta_{\rm stat}$($\%$) & $\delta_{\rm syst}$($\%$) & $\delta_{\rm tot}$($\%$) & ~~~NLO QCD~~~ & ~~~{\sc sherpa}~~~ & ~~~{\sc alpgen}~~~\\\hline
     20--30  & 24.3 &  9.6 $\times 10^{-2}$ & 2.4 & 17.8  & 18.0 &  6.5$\times 10^{-2}$  &  3.9$\times 10^{-2}$ & 3.9$\times 10^{-2}$            \\\hline
     30--40  & 34.3 &  4.0 $\times 10^{-2}$ & 2.9 & 13.6  & 13.9 &  3.0$\times 10^{-2}$  &  2.0$\times 10^{-2}$ & 2.0$\times 10^{-2}$            \\\hline
     40--50  & 44.3 &  2.5 $\times 10^{-2}$ & 3.6 & 14.4  & 14.8 &  1.6$\times 10^{-2}$  &  1.1$\times 10^{-2}$ & 1.1$\times 10^{-2}$           \\\hline
     50--70  & 57.2 &  1.2 $\times 10^{-2}$ & 3.4 & 15.2  & 15.6 &  7.4$\times 10^{-3}$  &  5.5$\times 10^{-3}$ & 5.2$\times 10^{-3}$            \\\hline
     70--150 & 81.7 &  2.2 $\times 10^{-3}$ & 4.5 & 17.7  & 18.3 &  1.4$\times 10^{-3}$  &  1.0$\times 10^{-3}$ & 9.3$\times 10^{-4}$            \\\hline
\end{tabular}
\end{center}
\end{table*}
\begin{table*}
\begin{center}
\caption{The \wc production cross sections times $W \to \mu\nu$ branching fraction, ${\rm d}\sigma/{\rm d}p_T^{\rm jet}$, 
together with statistical uncertainties ($\delta_{\text{stat}}$) and  total systematic uncertainties ($\delta_{\text{syst}}$). 
The column $\delta_{\rm tot}$ shows total experimental uncertainty obtained by adding $\delta_{\text{stat}}$
and $\delta_{\text{syst}}$ in quadrature.
The last three columns show theoretical predictions obtained using NLO QCD with MSTW PDF set, and two MC event generators,
{\sc sherpa} and {\sc alpgen}.}
\label{tab:res_c}
\begin{tabular}{ccccccccc} \hline \hline
~$p_T^{\rm jet}$ bin~ & ~$\la p_T^{\rm jet}\ra$~ & \multicolumn{7}{c}{${\rm d}\sigma/{\rm d}p_T^{\rm jet}$ (pb/GeV) } \\\cline{3-9}
 (GeV) & (GeV) & Data & $\delta_{\rm stat}$($\%$) & $\delta_{\rm syst}$($\%$) & $\delta_{\rm tot}$($\%$) & ~~~NLO QCD~~~ & ~~~{\sc sherpa}~~~ & ~~~{\sc alpgen}~~~\\\hline
     20--30  & 24.2  & 4.1$\times 10^{-1}$  &  3.7  &  17.0  &  17.4 &  4.1$\times 10^{-1}$  &  2.1$\times 10^{-1}$  & 2.4$\times 10^{-1}$   \\\hline
     30--40  & 34.2  & 2.6$\times 10^{-1}$  &  4.6  &  11.0  &  11.9 &  1.8$\times 10^{-1}$  &  9.2$\times 10^{-2}$  & 1.1$\times 10^{-1}$   \\\hline
     40--50  & 44.2  & 1.5$\times 10^{-1}$  &  5.8  &  11.9  &  13.2 &  9.2$\times 10^{-2}$  &  4.6$\times 10^{-2}$  & 5.9$\times 10^{-2}$   \\\hline
     50--70  & 57.0  & 8.4$\times 10^{-2}$  &  5.3  &  12.1  &  13.2 &  3.9$\times 10^{-2}$  &  2.0$\times 10^{-2}$  & 2.6$\times 10^{-2}$   \\\hline
     70--150 & 80.7  & 1.3$\times 10^{-2}$  &  6.9  &  15.6  &  17.1 &  6.1$\times 10^{-3}$  &  3.1$\times 10^{-3}$  & 3.8$\times 10^{-3}$    \\\hline
\end{tabular}
\end{center}
\end{table*}

%
%
%%% WHF data / theory ratio
%
\begin{figure}
\vskip -3mm
\includegraphics[width=0.45\textwidth]{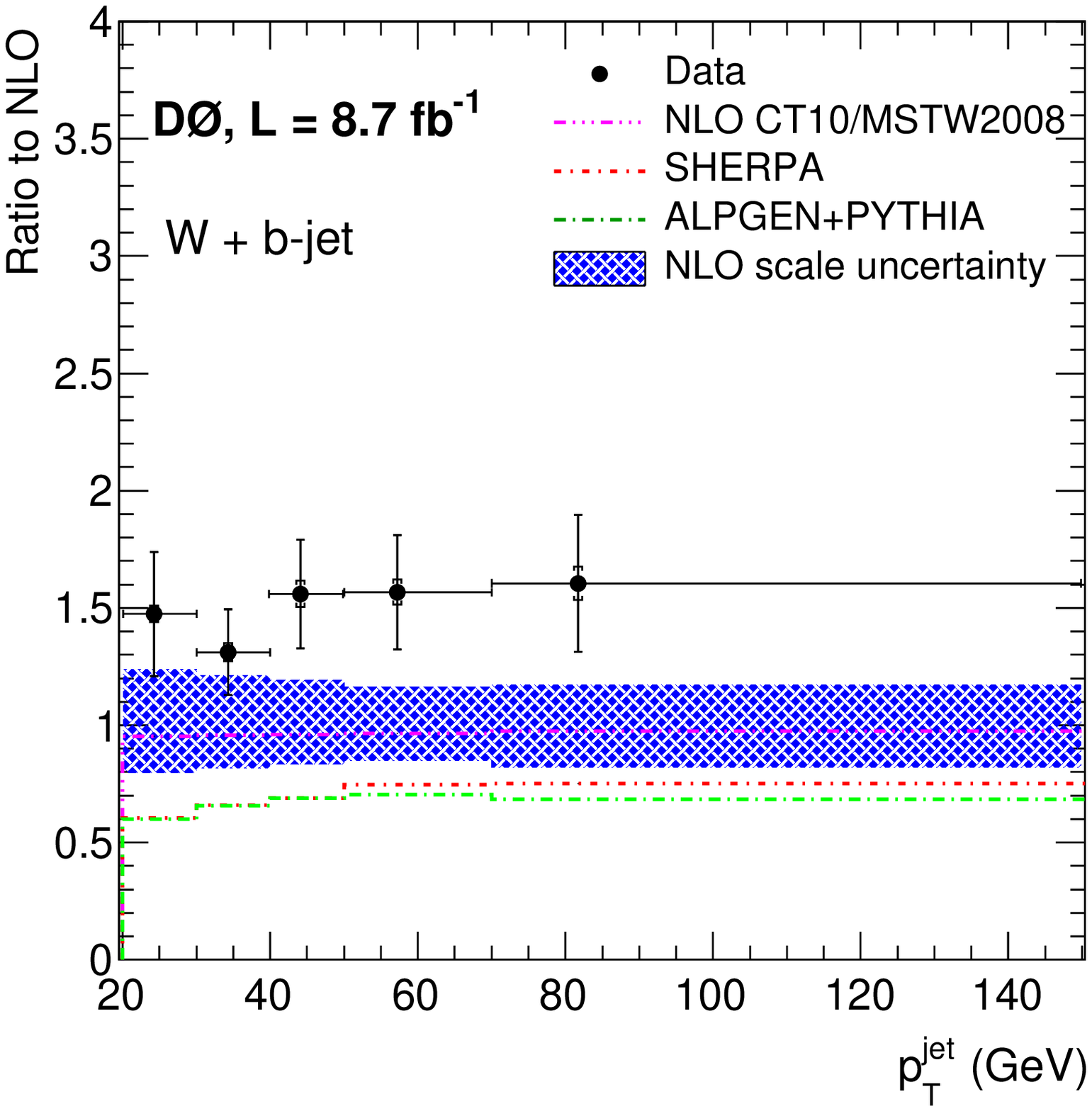}
\caption{(Color online)
The ratio of \wb production cross sections to NLO predictions with the MSTW2008 PDF set \cite{mstw} for data and theoretical predictions.
The uncertainties on the data include both statistical (inner error bar) and total uncertainties (full error bar).
Also shown are the uncertainties on the theoretical QCD scales.
The ratio of NLO QCD predictions with CT10 \cite{cteq}  to those obtained with MSTW2008
as well as the predictions given by {\sc sherpa}~\cite{sherpa} and {\sc alpgen}~\cite{alpgen} are also presented. }
 \label{fig:xsce_MSTW_wb}
\end{figure}

\begin{figure}
\includegraphics[width=0.45\textwidth]{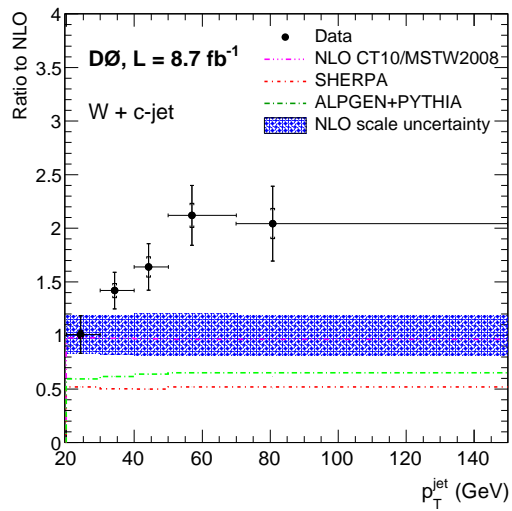}
\caption{(Color online)
The ratio of \wc production cross sections to NLO predictions with the  MSTW2008 PDF set \cite{mstw} for data and theoretical predictions.
The uncertainties on the data include both statistical (inner error bar) and total uncertainties (full error bar).
Also shown are the uncertainties on the theoretical QCD scales.
The ratio of NLO QCD predictions with CT10 \cite{cteq}  to those obtained with MSTW2008
as well as the predictions given by {\sc sherpa}~\cite{sherpa} and {\sc alpgen}~\cite{alpgen} are also presented.}
\label{fig:xsce_MSTW_wc}
\end{figure}

The NLO predictions are corrected for non-perturbative effects
such as parton-to-hadron fragmentation and multiple parton interactions.
The latter are evaluated using {\sc sherpa} and {\sc pythia} MC samples generated using their default
settings~\cite{sherpa,pythia}.
The overall corrections vary within a factor of $0.80-1.1$ with an uncertainty of $\lesssim 5\%$ assigned to account for the
difference between the two MC generators.
The ratios of data over the NLO QCD calculations and of the various theoretical predictions to the NLO
QCD calculations are presented in Figs.~\ref{fig:xsce_MSTW_wb} and \ref{fig:xsce_MSTW_wc}.
The measured \wb cross sections are systematically above the NLO QCD predictions for all jet $p_T$ bins.
The \wc data agree with the NLO QCD predictions at small $p_T$ but disagree at higher $p_T$ 
as the contribution from  $q \bar{q}' \to W+g~(g\to c\bar{c})$ events increases.

\begin{figure}[!h]
\vskip -3mm
\includegraphics[width=0.45\textwidth]{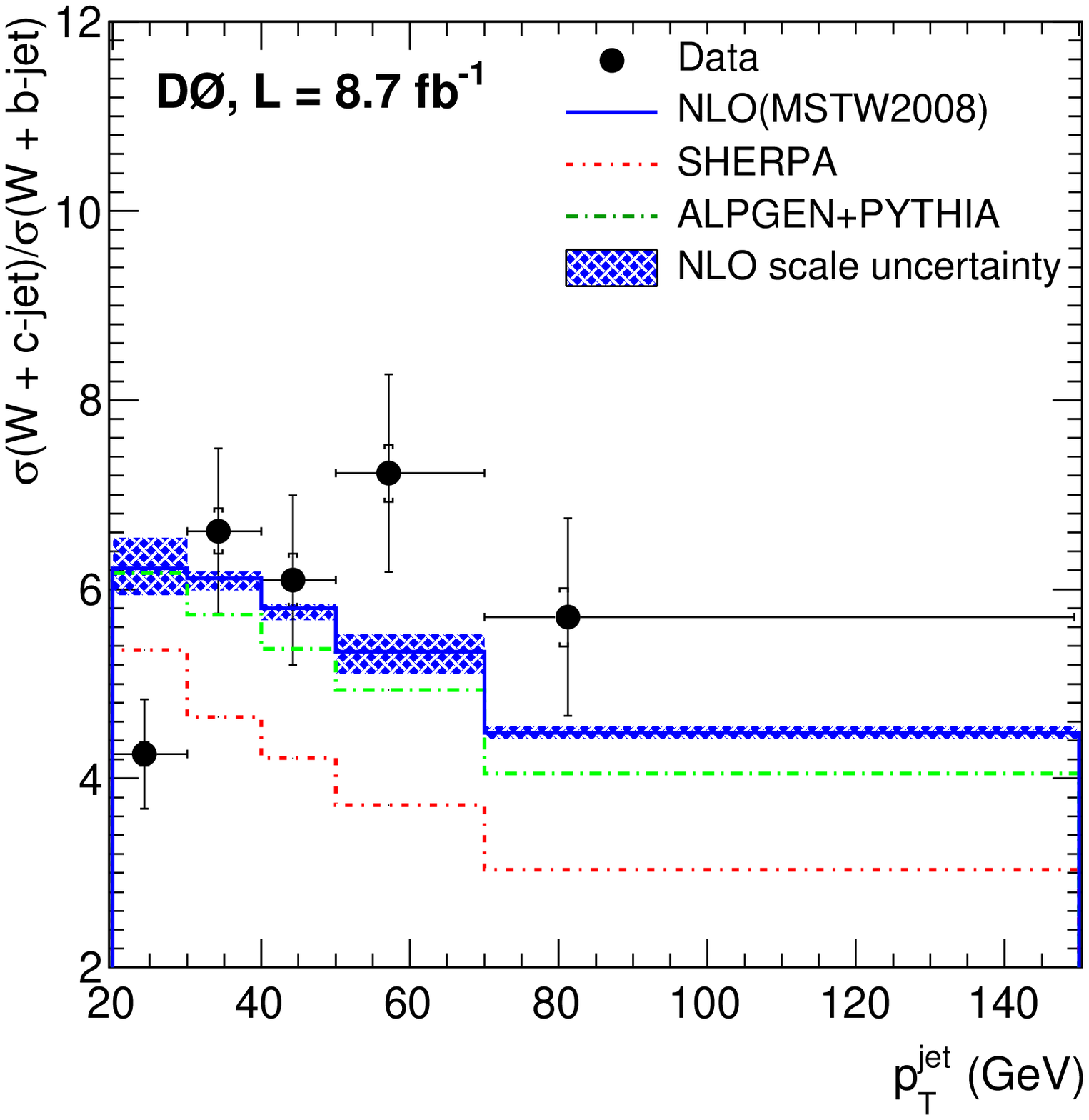}
\caption{The ratio of the \wc to \wb production cross sections for data and theory as a function of jet $p_T$.  The uncertainties on the points in
data include both statistical (inner line) and the full uncertainties (the entire error bar).
% (the entire line). 
Predictions given by NLO QCD with the  MSTW2008 PDF set\cite{mstw}, {\sc sherpa} \cite{sherpa} and {\sc alpgen} \cite{alpgen} are also shown. }
\label{fig:xsce_ratio_wc_wb}
\end{figure}

% Ratio
In addition to measuring the \wb and \wc cross-sections,
we  calculate the ratio $\sigma(W+c)/\sigma(W+b)$ in jet $p_T$ bins.
In this ratio, many experimental systematic uncertainties cancel.
Also, theory predictions of the ratio are less sensitive to the scale uncertainties
and effects from missing higher-order terms that impact the normalizations of the cross sections.
The remaining uncertainties are caused by largely anti-correlated uncertainties
coming from the fitting of $c$-jet and $b$-jet $D_{\rm MJL}$ templates to data,
and by other uncertainties on the $b$- and $c$-jet fractions discussed above.
Experimental results as well as theoretical predictions for the ratios are presented in Table \ref{tab:res_c_b_ratio} and Fig.~\ref{fig:xsce_ratio_wc_wb}.
The systematic uncertainties on the ratio vary within ($11-17$)\%.
Theoretical scale uncertainties, estimated by
varying the renormalization and factorization scales by a factor of two 
in the same way for the $\sigma(W+b)$ and $\sigma(W+c)$  predictions, are also significantly reduced.
Specifically, residual scale uncertainties are typically ($0.5-4.6$)\% for NLO QCD,
which indicates a much smaller dependence of the ratio on the higher-order corrections.
% The  ratio $\sigma(W+c)/\sigma(W+b)$ is consistent with theoretical predictions. 
The ratio $\sigma(W+c)/\sigma(W+b)$ for $p_T^{jet}> 30$ GeV is reasonably consistent with theoretical predictions except for {\sc sherpa}.
%
%%% Wc/Wb ratio
%

\begin{table*}
\begin{center}
\caption{
The $\sigma(W+c)/\sigma(W+b)$ cross section ratio in bins of $c(b)$-jet $p_T$ 
together with statistical uncertainties ($\delta_{\text{stat}}$),  total systematic uncertainties ($\delta_{\text{syst}}$). 
The column $\delta_{\rm tot}$ shows total experimental uncertainty obtained by adding $\delta_{\text{stat}}$
and $\delta_{\text{syst}}$ in quadrature.
The last three columns show theoretical predictions obtained with the NLO QCD using MSTW2008 PDF set, and two MC event generators,
{\sc sherpa} and {\sc alpgen}.
}
\label{tab:res_c_b_ratio}
\begin{tabular}{ccccccccc} \hline \hline
~$p_T^{\rm jet}$ bin~ & ~$\la p_T^{\rm jet}\ra$~ & \multicolumn{7}{c}{Ratio~$\sigma(W+c)/\sigma(W+b)$ } \\\cline{3-9}
 (GeV) & (GeV) & Data & $\delta_{\rm stat}$($\%$) & $\delta_{\rm syst}$($\%$) & $\delta_{\rm tot}$($\%$) & ~~~NLO QCD~~~ & ~~~{\sc sherpa}~~~ & ~~~{\sc alpgen}~~~\\\hline
     20--30 & 24.3 & 4.3  & 2.9 & 13.3 & 13.6 & 6.2  &  5.4  &   6.2          \\\hline
     30--40 & 34.3 & 6.6  & 3.6 & 12.7 & 13.2 & 6.1  &  4.7  &   5.7          \\\hline
     40--50 & 44.3 & 6.1  & 4.6 & 13.9 & 14.7 & 5.8  &  4.2  &   5.4        \\\hline
     50--70 & 57.1 & 7.2  & 4.2 & 13.8 & 14.4 & 5.3  &  3.7  &   4.9      \\\hline
     70--150& 81.2 & 5.7  & 5.4 & 17.5 & 18.3 & 4.5  &  3.0  &   4.1       \\\hline
\end{tabular}
\end{center}
\end{table*}

% Summary
%The small experimental uncertainty should allow to further constrain theoretical predictions. 
In summary, we have performed 
%a
the first measurement of the differential cross section as a function of $p_T^{\rm jet}$ for the 
%inclusive
 \wb and \wc final states with $W \to \mu\nu$ decay
at $\sqrt{s}=1.96$~TeV, in a restricted phase space of $p_T^{\mu}>20$~GeV, 
$|\eta^\mu|<1.7$, $p_T^{\nu}>25$~GeV and with $b(c)$ jets with the $p_T$ range
% limited to 
$20<p_T^{\rm jet}<150$~GeV and $|\eta^{\rm jet}| < 1.5$.  
These are the first measurements of $W+b/c$ cross sections that are sensitive to the gluon splitting processes.
The measured \wb cross section is higher than the predictions 
in all $p_T$ bins
% for all jet $p_T$ range, indicating 
and is suggestive of 
missing higher order corrections.
%  for all $p_T$ bins.
The measured \wc cross section agrees with NLO prediction for the low $p_T^{\rm jet}$ (20--30 GeV), but disagrees towards high $p_T^{\rm jet}$.
The disagreement may be due to missing higher order corrections and an underestimated contribution from gluon splitting $g\to c\bar{c}$ 
also observed earlier at LEP~\cite{LEP}, LHCb \cite{LHCb}, ATLAS \cite{Atlas} and D0 experiments \cite{gamma_c,Zc}, and/or
possible enhancement in the strange quark PDF as suggested by CHORUS \cite{chorus_strange}, CMS \cite{CMS_2014} and ATLAS \cite{atlas_strange}  data
according to a recent PDF fit performed by ABKM group \cite{Alekhin_strange}.

%Such a discrepency may be caused by an underestimated contribution from gluon splitting  $g\rightarrow c\bar{c}$~\cite{gcc1, gcc2, gcc3, gcc4, gcc5} 
%in the annihilation process or by contribution from intrinsic charm.

We thank John Campbell for useful discussions and predictions with \textsc{mcfm}.
% \input acknowledgement.tex   % input acknowledgement
% % \RequirePackage{lineno}     % to get line numbering
% 
% %\documentclass[aps,preprint,showpacs,groupedaddress]{revtex4}  % for double-spaced preprint
% %\documentclass[prd,amsmath,amssymb] {revtex4}                  % conference format
% %\documentclass[aps,showpacs,groupedaddress]{revtex4}  % single spaced preprint                                                    
% 
% \documentclass[12pt,titlepage]{report}
% \usepackage{graphicx}
% \usepackage{placeins}
% \usepackage{amssymb}
% \usepackage[dvips,usenames]{color}
% \usepackage{epsfig}
% \usepackage{proposal}
% 
% 
% 
% \def\etal{{\sl et al.}}
% %\newcommand{\met}{\mbox{\ensuremath{\slash\kern-.7emE_{T}}}}
% 
% \begin{document}
% \input epsf.tex    %<-If you need EPS figures to be
%                    %  called in {figure} environment for PC
% %\input epsf.def   %<-If you need EPS figures to be
%                    %  called in {figure} environment for Macintosh
% 
% %\input psfig.sty
% 
% 
% 
%  
% \noindent
% {\bf Acknowledgement paragraph with agency abbreviations of Oct. 15, 2014 for non-APS journals} 
%\vskip 5mm
We thank the staffs at Fermilab and collaborating institutions,
and acknowledge support from the
DOE and NSF (USA);
CEA and CNRS/IN2P3 (France);
MON, NRC KI, and RFBR (Russia);
CNPq and FAPERJ (Brazil);
DAE and DST (India);
COLCIENCIAS (Colombia);
CONACYT (Mexico);
NRF (Korea);
FOM (The Netherlands);
STFC and The Royal Society (UK);
MSMT (Czech Republic); 
BMBF and DFG (Germany);
SFI (Ireland);
Swedish Research Council (Sweden);
CAS and CNSF (China);
and
MESU (Ukraine).

%

% \end{document}


\begin{thebibliography}{99}
\bibitem{coord} 
We use a standard right-handed coordinate system. The nominal collision point is the center of the detector with coordinates $(0,0,0)$. 
The direction of the proton beam is the $+z$ axis. The $+x$ axis is horizontal, pointing away from the center of the Tevatron ring. 
The $+y$ axis points vertically upwards. The polar angle, $\theta$, is defined such that $\theta = 0 $ is the $+z$ direction. 
The rapidity is defined as $y = -\ln \left[ (E+p_z)/(E-p_z)\right]$, where $E$ is the energy and $p_z$ is the momentum component along 
the proton beam direction. Pseudorapidity is defined as $\eta = - \ln(\tan \frac{\theta}{2})$. 
$\phi$ is defined as the azimuthal angle in the plane transverse to the proton beam direction.
\bibitem{mt}
The measured properties of W boson candidates are limited by their transverse energy and transverse mass, defined as $M_T=\sqrt{(\not\!{p}_T+p^l_T)^2-(\not\!{p}_x+p^l_x)^2-(\not\!{p}_y+p^l_y)^2}$, where $\not\!{p}_T$ is the magnitude of the missing transverse energy vector, $p^l_T$ is the transverse momentum of the lepton and $p^l_x$ and $p^l_y$ ($\not\!{p}^l_x$ and $\not\!{p}^l_y$) are the magnitude of the $x$ and $y$ components of the lepton's momentum (missing transverse energy), respectively.
\end{thebibliography}
\end{document}